\documentclass[aps,twocolumn,pra,superscriptaddress,showpacs,tightenlines]{revtex4}
\usepackage{amssymb}
\usepackage{amsmath}
\usepackage{graphicx}
\usepackage{epsfig}
\usepackage{subfigure}
\usepackage{amsfonts}
\usepackage{CJK}

\begin{document}

\title{Quantum Zeno switch for single-photon coherent transport}
\author{Lan Zhou}
\affiliation{Department of Physics, Hunan Normal University,
Changsha 410081, China} \affiliation{Advanced Science Institute, The
Institute of Physical and Chemical Research (RIKEN), Wako-shi
351-0198, Japan}
\author{S. Yang}
\affiliation{Institute of Theoretical Physics, The Chinese Academy
of Sciences, Beijing, 100080, China}
\author{Yu-xi Liu}
\affiliation{Advanced Science Institute, The Institute of Physical
and Chemical Research (RIKEN), Wako-shi 351-0198, Japan}
\affiliation{CREST, Japan Science and Technology Agency (JST),
Kawaguchi, Saitama 332-0012, Japan}
\author{C. P. Sun}
\affiliation{Institute of Theoretical Physics, The Chinese Academy
of Sciences, Beijing, 100080, China} \affiliation{Advanced Science
Institute, The Institute of Physical and Chemical Research (RIKEN),
Wako-shi 351-0198, Japan}
\author{Franco Nori}
\affiliation{Advanced Science Institute, The Institute of Physical
and Chemical Research (RIKEN), Wako-shi 351-0198, Japan}
\affiliation{CREST, Japan Science and Technology Agency (JST),
Kawaguchi, Saitama 332-0012, Japan} \affiliation{Center for
Theoretical Physics, Physics Department, Center for the Study of
Complex Systems, The University of Michigan, Ann Arbor, Michigan
48109-1040, USA.}
\begin{abstract}
Using a dynamical quantum Zeno effect, we propose a general approach
to control the coupling between a two-level system (TLS) and its
surroundings, by modulating the energy level spacing of the TLS with
a high frequency signal. We show that the TLS--surroundings
interaction can be turned on or off when the ratio between the
amplitude and the frequency of the modulating field is adjusted to
be a zero of a Bessel function. The quantum Zeno effect of the TLS
can also be observed by the vanishing of the photon reflection at
these zeros. Based on these results, we propose a quantum switch to
control the transport of a single photon in a 1D waveguide. Our
analytical results agree well with numerical results using Floquet
theory.
\end{abstract}
\maketitle \narrowtext 
\narrowtext

\section{\label{Sec:1}Introduction}

The quantum Zeno effect (QZE)~\cite{BESJMP18} was once described as the
quantum version of the expression \textquotedblleft a watched pot never
boils\textquotedblright. Namely, an unstable particle, if observed
constantly, will not decay~\cite{KSPR412,HWAP258,FJPRA70,WXBPRA77,HMPRA77}.
This effect has been explained (e.g., Refs.~\cite{BESJMP18,KurN405,ZHLSPRA74}%
) in terms of wave-packet collapse: once a quantum measurement is performed,
the superposed quantum state is reduced to a single one of the allowed
eigenstates of the measured observable.

However, many authors (e.g., Refs.~\cite%
{PerAJP48,pripA170,BalPRA43,BBPRA44,SYLfp43,llyod,PDSPRA69}) explained the
QZE~\cite{winPRA41,FPRL87,SPRL97} \textit{without} invoking the wave
function collapse. This \textit{dynamical} QZE does not involve quantum
states collapsing in one eigenstate, and thus evolves in time. Indeed, Asher
Peres~\cite{PerAJP48} argued that the decay of an unstable quantum system
can be slowed down and even halted by its continual interaction with an
external system, without measurement or observation. This dynamical QZE is
sometimes called ``bang-bang'' control (e.g., in Refs.~\cite%
{SYLfp43,llyod,PDSPRA69}). Here, we apply the dynamic QZE to realize a
quantum switch.

The QZE can be used as a dynamical mechanism to suppress the
environment-induced decoherence of a quantum system~\cite{KurN405,PDSPRA69}.
This implies that the coupling between the system and its environment can be
controlled via the QZE. Here, we propose a quantum ``Zeno switch'' to turn
on or off the quantum coherent transport of particles along a
one-dimensional (1D) continuum.

A quantum switch~\cite{Sun1,fanpaper,ZGLSN} is regarded as an active device
to control the transport of particles or the transfer of quantum states. It
plays a similar role to the gate voltage, which turns on or off the
transmission of electrons, in conventional electronic circuits. The quantum
version of the switch means that the operations involve a single quantum. An
example is the single-photon transistor~\cite%
{Lukin-np,cavity,BRPRA74,cavity2}, where the transport of a single photon in
a 1D waveguide can be controlled by a two-level system (TLS) externally
driven by a classical field.

A discrete coordinate scattering approach~\cite{ZGLSN} was developed to
study a quantum switch for the coherent transport of a single photon along a
1D coupled-resonator waveguide (CRW). That quantum switch~\cite{ZGLSN} can
be realized by changing either the energy level spacing or the coupling
constants between the waveguide and a controller. Physical systems proposed
to realize single-photon switches include: (i) Superconducting transmission
line resonators coupled to a superconducting charge qubit~\cite{ZGLSN}; (ii)
A TLS coupled to a photonic crystal ``defect-cavity'' waveguide~\cite%
{fanpaper}. For the case (i), the controllability can be realized by
changing the energy level spacing of the charge qubit. However, it seems
difficult to control well the photon transport in (ii), because the
couplings and TLS parameters are fixed once the sample is fabricated.

\begin{figure}[ptb]
\includegraphics[bb=45 378 493 558, width=8 cm, clip]{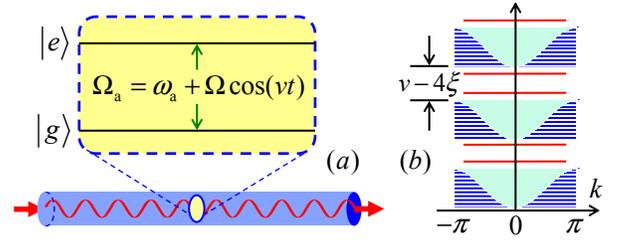}
\caption{(color online) (a) Schematic diagram of a quantum switch: the
coupled resonator waveguide is coupled to a TLS with an energy level spacing
modulated by $\Omega_{a}(t)$. The Rabi oscillation can appear in the
effective multi-band structure of frequency shown in (b), with bound states
(in red) in the gap.}
\label{setup}
\end{figure}

In this paper , for a single-photon propagating in a 1D CRW, a dynamical QZE
switch is proposed with a tunable effective coupling to a TLS. This coupling
is controlled by an applied frequency-modulated electromagnetic field. The
photon can be transmitted either directly through the continuum, or
indirectly, via a discrete energy level provided by the TLS. These two
channels interfere with each other. Their destructive quantum interference
prevents the photon transport while the constructive one leads to a total
transmission ~\cite{SAFPRA75}. The dynamic Zeno effect of the TLS could be
actively controlled by a frequency modulation, to switch between the two
quantum interference channels. When the dynamic Zeno effect slows down the
transitions in the TLS, the photon transport appears only when the TLS is in
its ground state, no destructive quantum interference occurs. Then,
effectively, the TLS and the continuum are decoupled.

\section{\label{Sec:2} Decoupling mechanism using dynamical QZE.}

We generally consider a quantum system $S$ with a characterized frequency $%
\omega _{c}$ and free Hamiltonian $H_{e}$, coupled to a TLS, with its ground
$|g\rangle $ and excited $|e\rangle $ states, and energy level spacing $%
\omega _{\mathrm{a}}$ (with $\hbar =1$). We assume that a periodically
modulated field is applied to TLS so that
\begin{equation}
\omega _{\mathrm{a}}\rightarrow \Omega _{\mathrm{a}}(t)=\omega _{\mathrm{a}%
}+\Omega \cos (\nu t),
\end{equation}%
where $\Omega $ is the amplitude of the modulation with frequency $\nu $.
Hamiltonian
\begin{equation}
H=\Omega _{\mathrm{a}}(t)|e\rangle \langle e|+G|e\rangle \langle g|+G^{\ast
}|g\rangle \langle e|+H_{e}
\end{equation}%
describes the quantum system $S$ and the TLS, as well as the coupling
between them. The coupling coefficient $G$ depends on $S$-variables so that
\begin{equation}
e^{iH_{e}t}Ge^{-iH_{e}t}=Ge^{-i\omega _{c}t}.
\end{equation}

In the interaction picture, the \textit{coupling} between $S$ and the TLS is
modeled by the Hamiltonian
\begin{equation}
H_{\mathrm{I}}=G\sum_{n=-\infty }^{+\infty }J_{n}\left( \frac{\Omega }{\nu }%
\right) e^{i(n\nu -\Delta )t}|e\rangle \langle g|+\text{H.c.}\,,
\label{qdz-01}
\end{equation}%
where the detuning $\Delta =\omega _{c}-\omega _{\mathrm{a}}$. In Eq.~(\ref%
{qdz-01}), we have used~\cite{SAFPRA75} the Fourier-Bessel series identity,
\begin{equation*}
e^{ix\sin \gamma }=\sum_{n}J_{n}(x)e^{in\gamma },
\end{equation*}%
with the $n$th Bessel function $J_{n}(x)$ of the first kind.

For a fast modulation with large $\nu$, the lowest frequency terms in Eq.~(%
\ref{qdz-01}) dominate the dynamical evolution. These terms are determined
by the near-resonant and resonant condition $(n\nu-\Delta)\simeq0$; here $n=%
\left[ \Delta/\nu \right] $ is the integer nearest to $\Delta/\nu$. To the
lowest frequency term of the index $n=\left[ \Delta/\nu \right] $ , Eq.~(\ref%
{qdz-01}) is approximated as
\begin{equation}
H_{\mathrm{I}}\approx G\;J_{[\Delta/\nu]}\! \left( \frac{\Omega}{\nu}\right)
e^{(i\left[ \Delta/\nu \right] \nu-\Delta)t}|e\rangle \langle g|+\text{H.c.}%
\,.  \label{dz-03}
\end{equation}
For a very fast modulation of frequency $\nu$ (i.e., $\Delta/\nu \sim0$) and
a large $\Omega$, we have%
\begin{equation*}
H_{I}\propto J_{0}(\Omega/\nu).
\end{equation*}
The values of the decoupling points (i.e., $\Omega/\nu \cong2.40,5.52,...$)
are just the zeros of $J_{0}(x)$. Equation~(\ref{dz-03}) clearly shows that
the \textit{effective interaction vanishes} when the ratio $\Omega/\nu$
(=amplitude/frequency of the driving field) is a zero of the Bessel function
$J_{[\Delta/\nu]}\left( x\right) $, therefore \textit{decoupling} $S$ from
the TLS. At these zeros, the dynamical QZE occurs; namely, the TLS initially
prepared in its excited state will remain there and will not decay to the
ground state.

The above arguments to realize the QZE dynamics does not depends on the
concrete form of the quantum system $S.$ Thus its based idea can be
generally used for the various decoupling schemes.

Notice that the Bessel functions $J_{n}\left( x\right) $ look roughly like
oscillating sine or cosine functions, and have an infinite number of zeros.
The decays of the Bessel functions are proportional to $1/\sqrt{x}$. Except
asymptotically large $x$, the roots of the Bessel functions are not periodic.

\section{\label{Sec:3}Single photon quantum switch in a 1D waveguide.}

Based on the above dynamical QZE mechanism, we now propose a quantum device,
which behaves as a switch to control the incident photon transport in a CRW
made of a periodic array of identical coupled resonators. The main
difference between this device and the one in Ref.~\cite{ZGLSN} is that here
the TLS energy level spacing is now modulated by a periodic field, allowing
us to use the remarkable dynamic QZE in a very unusual manner. The 1D CRW is
schematically shown in Fig.~\ref{setup}(a), where a TLS is embedded in one
of the resonators and is modulated with the external periodic forcing of
amplitude $\Omega $ and frequency $\nu $, i.e. $\omega _{\mathrm{a}%
}\rightarrow \Omega _{\mathrm{a}}(t)$. Let $a_{j}^{\dagger }$ ($j=-\infty
,\cdots ,\infty $) be the creation operator of the $j$th single mode cavity,
all with the same frequency $\omega $. The Hamiltonian of the CRW reads
\begin{equation}
H_{\mathrm{C}}=\omega _{c}\sum_{j}a_{j}^{\dagger }a_{j}-\xi \sum_{j}\left(
a_{j}^{\dagger }a_{j+1}+\mathrm{H.c.}\right) \,,  \label{lps-01}
\end{equation}%
with the inter-cavity coupling constant $\xi $, which describes the photon
moving from one cavity to another. For convenience, we take the $0$th cavity
as the coordinate-axis origin and also assume that the TLS is located in
this $0$th cavity. Under the rotating wave approximation, the interaction
between the $0$th cavity field and the TLS is described by a Jaynes-Cummings
Hamiltonian
\begin{equation}
H_{\mathrm{I}}=\Omega _{\mathrm{a}}(t)\left\vert e\right\rangle \left\langle
e\right\vert +g\left( a_{0}^{\dag }\left\vert g\right\rangle \left\langle
e\right\vert +\left\vert e\right\rangle \left\langle g\right\vert
a_{0}\right) ,  \label{lps-02}
\end{equation}%
with coupling strength $g$ and modulated transition frequency $\Omega _{%
\mathrm{a}}$ of the TLS. By employing the Fourier transformation
\begin{equation}
a_{j}=\frac{1}{\sqrt{N}}\sum_{k}e^{ikj}a_{k},  \label{lps-02a}
\end{equation}
the second term in Eq.~(\ref{lps-02}) and $H_{C}$ can be rewritten
in $k$-space. In the rotating frame with respect to $H_{C}+\Omega
_{\mathrm{a}}(t)\left\vert e\right\rangle \left\langle
e\right\vert$, the interaction Hamiltonian reads
\begin{equation}
H_{\mathrm{II}}(t)=\frac{g}{\sqrt{N}}\!\!\sum_{k;n}\!J_{n}\!\left( \frac{%
\Omega }{\nu }\right) \!\!\left[ a_{k}^{\dag }\sigma _{-}e^{i\left(
\Delta -\epsilon _{k}-n\upsilon \right) t}+\text{H.c.}\right] ,
\label{eq:5}
\end{equation}%
where we have used the Fourier-Bessel series identity. Here, the
dispersion relation $\epsilon _{k}=2\xi \cos k$ describes an energy
band of width $4\xi $ (the lattice constant $l$ is assumed to be
unity), $\sigma _{-}=|g\rangle\langle e|$. The $H_{\mathrm{II}}(t)$
in Eq.~(\ref{eq:5}) effectively describes a multi-band Rabi
oscillation.

In this work we are mainly interested in the investigation of the
transmission and localization of single photon in high-frequency
regimes, corresponding to $\Delta ,\xi \ll \nu $. For small $\xi $,
compared with the modulation frequency $\nu $, the gaps are large
and there is no energy band overlap [the band structure is
illustrated in Fig.~\ref{setup}(b)]. Otherwise there exists a
complex quantum dynamics with quantum chaos. In the high-frequency
regime, the fast modulation in $\Omega _{\mathrm{a}}(t)$ implies the
resonance condition $n=0$. Therefore, the dynamic of the CRW+TLS
is described by the effective Hamiltonian%
\begin{equation*}
H=H_{C}+\omega _{a}\left\vert e\right\rangle \left\langle
e\right\vert +gJ_{0}\left(\frac{\Omega}{\nu}\right)\!\left(
a_{0}^{\dag }\left\vert g\right\rangle \left\langle e\right\vert
+\left\vert e\right\rangle \left\langle g\right\vert a_{0}\right) .
\end{equation*}%
In the one excitation subspace, the wavefunction at arbitrary time
\begin{equation*}
\left\vert \phi \left( t\right) \right\rangle =\sum_{j}u_{j}\!\left(
t\right) e^{-i\omega _{a}t}\left\vert 1_{j}\,g\right\rangle +u_{e}\!\left(
t\right) e^{-i\omega _{a}t}\left\vert 0\,e\right\rangle
\end{equation*}%
is a superposition of the photon at the jth cavity with atom in the
ground state and no photon in all cavities with atom in the excited
state. The equations for the excited state amplitude $u_{e}(t)$ and
the amplitudes $u_{j}(t)$ of single photon states derived from the
Schroedinger equation with Hamiltonian $H$ are given by
\begin{subequations}
\label{ampEq}
\begin{align}
i\dot{u}_{j}\!\left( t\right) & =\Delta u_{j}\!\left( t\right) -\xi \left[
u_{j-1}\!\left( t\right) +u_{j+1}\!\left( t\right) \right] +Gu_{e}\!\left(
t\right) \!\delta _{j0}. \\
i\dot{u}_{e}\!\left( t\right) & \simeq G\,u_{0}\!\left( t\right)
\end{align}%
where the overdot indicates the derivative with respect to time. These
amplitude equations (\ref{ampEq}) show that the interaction between the
single photon and the TLS is characterized by the effective \textit{coupling}
constant
\end{subequations}
\begin{equation*}
G=g\,J_{0}\left( \Omega /\nu \right) .
\end{equation*}%
For zeros of zeros of $J_{0}\left( \Omega /\nu \right) $ , i.e.,%
\begin{equation*}
\Omega \cong 2.40\nu ,5.52\nu ,...
\end{equation*}%
the high frequency modulation $\Omega _{\mathrm{a}}(t)$ leads to the
decoupling between the TLS and the CRW, and thus photons in the CRW will
propagate freely without feeling the influence from the localized TLS.
Therefore, this modulated TLS acts as a quantum \textquotedblleft Zeno
switch\textquotedblright .

\section{Floquet-theory-based numerical simulation of the dynamical QZE}

Let us further study the QZE decoupling mechanism and the Zeno switch by
numerically solving the time-dependent Hamiltonian $H_{\mathrm{C}}+H_{%
\mathrm{I}}$. Note the total Hamiltonian is periodic in time, with period $%
2\pi/\nu$. Using Floquet theory, we define
\begin{equation}
H_{\mathrm{F}}=H_{\mathrm{C}}+H_{\mathrm{I}}-i\partial_{t}
\end{equation}
as in, e.g., Ref.~\cite{Floquet}. Each eigenstate $\left \vert
f_{n}(t)\right \rangle $, with eigenvalue $\varepsilon_{n}$, of $H_{F}$ can
be used to define an eigenstate
\begin{equation}
\left \vert F_{n}\right \rangle =\exp(-i\varepsilon_{n}t)\left \vert
f_{n}\right \rangle
\end{equation}
of $H_{\mathrm{F}}$ with zero eigenvalue. Then the superposition of these
eigenstates
\begin{equation}
\left \vert \Phi(t)\right \rangle =\sum_{n}c_{n}\left \vert F_{n}\right
\rangle
\end{equation}
determines the general solution of the time-dependent Schr\"{o}dinger
equation corresponding to $H_{\mathrm{C}}+H_{\mathrm{I}}$. With an initial
state $\left \vert \Phi(0)\right \rangle $, the coefficients $c_{n}$ are
given by
\begin{equation*}
c_{n}=\langle f_{n}(0)\left \vert \Phi(0)\right \rangle .
\end{equation*}
Our task now is to diagonalize $H_{\mathrm{F}}$ in a discrete Hilbert space
with the spatio-temporal basis vectors
\begin{equation}
\left \vert j;m_{t}\right \rangle =\left \vert j\right \rangle \otimes \left
\vert m_{t}\right \rangle ,
\end{equation}
where $m_{t}$ is the discrete temporal coordinate of the time $t$,
satisfying
\begin{equation}
\left \langle t\right \vert \left. m_{t}\right \rangle =\exp \left( i\nu
m_{t}t\right) /\sqrt{T},
\end{equation}
with a large $T$; $\left \vert j\right \rangle $ is the eigenstate of the
discrete space coordinate. On this time-varying basis $\{|j,m_{t}\rangle \}$%
, the matrix elements of $H_{F}$ are independent of time.

\begin{figure}[ptb]
\includegraphics[bb=13 213 586 773, width=7 cm, clip]{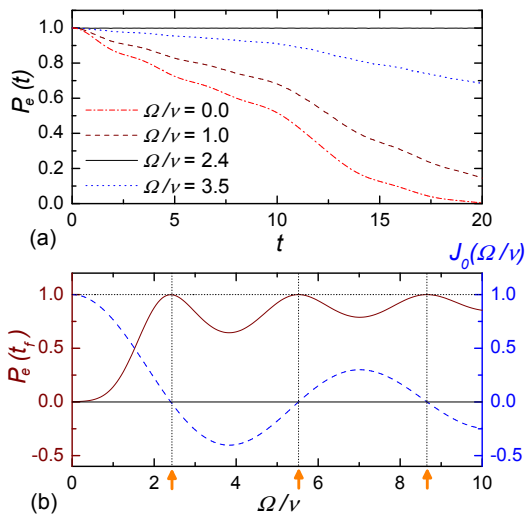}
\caption{(color online) (a) Time dependence of the probability $P_{e}(t)$
for the TLS at its excited state with different $\Omega/\protect\nu$, with
parameters: $\Delta=0$, $\protect\xi=1.0$, $g=0.25$, $\protect\nu=10$ and $%
L=41$ resonators; $t$ is in units of $1/\protect\xi$. Note that the excited
state does not decay (i.e., QZE) when $\Omega/\protect\nu=2.4$. (b) $%
P_{e}(t_{f})$ (wine solid line) versus $\Omega/\protect\nu$ is plotted for a
certain time $t_{f}=20$ using the same data as in (a). Obviously, the points
for $P_{e}(t_{f})=1$ (corresponding to no decay for the TLS) agree well with
the zeros (orange vertical arrows) of $J_{0}(\Omega/\protect\nu)$ (blue
dashed line).}
\label{fig:qdz-2}
\end{figure}

For the dynamic QZE in this paper, the initial state $\left \vert
\Phi(0)\right \rangle =|0,e;n_{t}=0\rangle$ evolves to $\left \vert
\Phi(t)\right \rangle $. To keep the numerical accuracy and save computing
time we divide the total time $t$ into $N$ small time intervals $\tau$, $%
t=N\tau$. In each interval, the system is evolved from $m\tau$ to $(m+1)\tau$%
. The final state of this interval becomes the initial state of the next
interval. For a sufficiently small value of $\tau$, an accurate wave
function can be obtained, even when not too many $\left \{ \left \vert
m_{t}\right \rangle \right \} $ are used to represent $H_{F}$~\cite{Floquet}.

Figure~\ref{fig:qdz-2}(a) shows the probability $P_{e}(t)$ for the TLS to be
in the excited state. When $\Omega/\nu$ is tuned to the vicinity of the
zeros of $J_{0}(x)$ (e.g., $\Omega/\nu=2.40$), the initially excited TLS
does not decay. Otherwise (e.g., $\Omega/\nu=1.0$ and $3.5$), the
probability $P_{e}(t)$ in the excited state decreases with time. Figure~\ref%
{fig:qdz-2}(b) shows how $P_{e}(t_{f})$ varies with the ratio $\Omega/\nu$
at the instant $t_{f}=20$ for $J_{0}(\Omega/\nu)$. It clearly demonstrates
that the zeros of $J_{0}(\Omega/\nu)$ correspond to $P_{e}(t_{f})=1$, i.e.,
the TLS does not decay. Thus the numerical results agree well with our
analytical results on the dynamical QZE.

\section{Quantum Zeno Dynamics}

In this section we give the above mentioned dynamic QZE a more standard
approach in current literatures \cite{KofPRL87}, which confirms the
suitability for using the idea of QZE.

Within the subspace spanned by $\left \vert e0\right \rangle $ and $
\left \vert gj\right \rangle $, if the TLS is prepared initially in
the excited state, the evolution of the TLS-CRW system will lead to
a superposition of states, corresponding to the TLS initially
excited and no photons in the CRW and the TLS in the ground state
and one photon in the CRW. The CRW forms a continuum. Therefore, the
interaction process between the TLS and the CRW can be described in
terms of a discrete level coupled to a continuum, and will lead to a
dissolution of the discrete state over an interval of width $R $,
which is the decay rate of the discrete state to the continuum. A
universal mechanism of quantum-mechanical decay control is usually
based on periodic coherent pulses. This control can yield either the
inhibition (Zeno effect) or the acceleration (anti-Zeno effect).

Following the analysis in Ref.\cite{KofPRL87}, we calculate the probability%
\begin{equation}
P_{e}\left( t\right) =\exp \left[ -R\left( t\right) Q\left( t\right) \right]
\end{equation}%
to reflects the decay law (for detailed caculation see the appedix), where
\begin{equation}
Q\left( t\right) =\int_{0}^{t}d\tau \left\vert V\left( \tau \right)
\right\vert ^{2}
\end{equation}%
is the effective interaction time with%
\begin{equation}
V\left( t\right) =\exp \left[ -i\frac{\Omega }{\upsilon }\sin \left(
\upsilon t\right) \right]
\end{equation}%
here. In the frequency domain, the effective decay rate
\begin{equation}
R\left( t\right) =\int_{-\infty }^{\infty }\tilde{\Phi}\left( \omega +\Delta
\right) F_{t}\left( \omega \right) d\omega
\end{equation}%
is the overlap of the normalized spectral modulation intensity
\begin{equation}
F_{t}\left( \omega \right) =\left\vert \int_{0}^{t}d\tau V\left( \tau
\right) \mathrm{e}^{-i\omega \tau }\right\vert /Q\left( t\right) ,
\end{equation}%
and the reservoir coupling spectrum $\tilde{\Phi}\left( \omega \right) $,
which is the Fourier transformation of the memory function or reservoir
response function
\begin{equation}
\Phi \left( t\right) =\frac{J^{2}}{N}\sum_{k}\mathrm{e}^{i2\xi t\cos k}.
\end{equation}%
Obviously, $V\left( t\right) $ is periodic with the period $\upsilon
^{-1}$, the Bessel function $J_{n}(A/\upsilon )$ of the first kind
becomes the Fourier components of $V\left( t\right) $. At time $t$
much larger than the period $\upsilon ^{-1}$, the decay rate reads
\begin{equation}
R\left( t\right) =t\sum_{n=-\infty }^{+\infty }J_{n}^{2}\left( \frac{\Omega
}{\upsilon }\right) \int_{-\infty }^{\infty }\tilde{\Phi}\left( \omega
\right) \mathrm{sinc}^{2}\frac{\left( \omega -\Delta -n\upsilon \right) t}{2}%
d\omega ,
\end{equation}%
where \textrm{sinc}$x=\sin x/x$. Figure \ref{fig:qdz-2} roughly
shows the dependence of the decay rate on the ratio of the driving
amplitude to the driving frequency. As we mentioned before, we are
interested in the high-frequency regime with parameters $\Delta ,\xi
\ll \nu $. Since the width of the reservoir spectrum is proportional
to $\xi$, the relation of these parameters imply that the modulation
frequency is much greater than the inverse correlation time of the
continuum. Therefore, $\tilde{\Phi}\left( \omega \right)$ does not
change significantly over the spectral intervals $\nu ^{-1}$. In
this case, we can make the approximation
\begin{equation}
R\left( t\right) =tJ_{0}^{2}\left( \frac{\Omega }{\upsilon }\right) \mathrm{%
sinc}^{2}\frac{\left( \omega -\Delta \right) t}{2}\int_{-\infty }^{\infty }%
\tilde{\Phi}\left( \omega \right) d\omega .
\end{equation}%
The above equation describes that the state decays into all the
channels of the reservoir. Since the effective decay rate is
averaged over all decay channels \cite{KurN405}, we work in the QZE
regime. Obviously, the decay rate without modulation is suppressed
by a factor $J_{0}^{2}\left( A/\upsilon \right) $ when a periodical
modulation is applied. When $t$ is much larger than both $\upsilon
^{-1}$ and an effective correlation time of the reservoir, one
obtains
\begin{equation}
R\left( t\right) =J_{0}^{2}\left( \frac{\Omega }{\upsilon }\right) \tilde{%
\Phi}\left( \Delta \right)
\end{equation}%
i.e. the extension of the golden rule rate to the case of a
time-dependent coupling \cite{KofPRL87}. Therefore, the atom remains
in its initial state when the ratio between the modulation amplitude
and the frequency meets the zeroes of the Bessel function. We also
note that when $A=0$, the $0$th Bessel function $ J_{0}\left(
0\right) =1$, which means the decay rate $R\left(
t\right)=\tilde{\Phi}\left( \Delta \right) $ without any modulation.

\section{\label{Sec:5}Scattered and localized photons with dynamical QZE}

We now study the photon localization and the measurement on the QZE via
photon scattering. The purpose for this study is twofold: (i) Utilize the
QZE-based mechanism to turn on or off the transmission of photons in the
CRW, so that an ideal single-photon transistor can be realized; (ii) Witness
the dynamical QZE in the TLS by measuring the single photon scattering. We
further consider the relation between photon localization and the appearance
of dynamic QZE.

Solutions to Eqs. (\ref{ampEq}) can be found in the form of either
localized states around the location of the TLS or as a
superposition of extended propagating Bloch waves incident,
reflected and transmitted by the TLS embedded in the CRW. It was
done by first Fourier transforming the equations of motion of
$u_{j}\left( t\right) $ and $u_{e}(t)$ in Eq.~(\ref{ampEq}) to
obtain their Fourier
transforms $U_{k}\left( j\right) $ and $U_{e}\left( j\right) $%
\begin{align}
E_{k}U_{k}\left( j\right) & =\Delta U_{k}\left( j\right) -\xi \left[
U_{k}\left( j-1\right) +U_{k}\left( j+1\right) \right] +GU_{e}\delta _{j0}
\notag \\
E_{k}U_{e}\left( j\right) & =GU_{k}\left( j\right) \delta _{j0}.
\end{align}%
Eliminating the amplitude $U_{e}\left( j\right) $ in the above
equation, we obtain the discrete-coordinate scattering equation
\begin{equation}
\lbrack V(E_{k})+\Delta -E_{k}]U_{k}\left( j\right) =\xi \lbrack U_{k}\left(
j-1\right) +U_{k}\left( j+1\right) ],  \label{eigeneq}
\end{equation}%
where
\begin{equation*}
V(E_{k})=g^{2}J_{0}^{2}(\Omega /\nu )/E_{k}
\end{equation*}%
is a resonant potential resulting from the second-order transition process
due to the coupling with the TLS. It behaves as an infinite $\delta $%
-potential on the resonance $E_{k}=0$. In the absence of the TLS, a solution
of Eqs. (\ref{ampEq}) has the form $U_{k}\left( j\right) =\exp (ikj)$, where
$k$ is the wave number of the Bloch waves. Therefore,
\begin{equation*}
E_{k}=\Delta -2\xi \cos k
\end{equation*}%
gives a band of width $4\xi $ with its minimum lies at $k=0$. Within the
band, the scattering wave function is assumed as
\begin{equation}
U_{k}\left( j\right) =\exp (ikj)+r\exp (-ikj),
\end{equation}%
for $j<0$, and
\begin{equation*}
U_{k}\left( j\right) =s\exp (ikj),
\end{equation*}%
for $j>0$, with the right- and left-moving waves $\exp (ikj)$ and $\exp
(-ikj)$. The boundary conditions
\begin{subequations}
\label{5boundc}
\begin{align}
U_{k}\left( 0^{+}\right) & =U_{k}\left( 0^{-}\right)  \\
\lbrack V(E_{k})+\Delta -E_{k}]U_{k}\left( 0\right) & =\xi \lbrack
U_{k}\left( -1\right) +U_{k}\left( 1\right) ]
\end{align}%
result in the reflection amplitude $r=s-1$ and transmission amplitude
\end{subequations}
\begin{equation}
s=\frac{\omega _{k}-\omega _{a}}{\omega _{k}-\omega
_{a}+ig^{2}J_{0}^{2}\left( \Omega /\nu \right) /v_{g}}.  \label{qzs0-tr}
\end{equation}%
Here,
\begin{equation*}
v_{g}=2\xi \sin k
\end{equation*}%
is the group velocity and
\begin{equation*}
\omega _{k}=\omega _{c}-2\xi \cos k
\end{equation*}%
is the incident energy of the single photon.

Equation~(\ref{qzs0-tr}) indicates that when $\omega _{a}$ is inside the
band $(\omega _{k}=\omega _{c}-2\xi \cos k)$, a dip down to zero will occur
in the transmission line-shape $|s|^{2}$. This zero transmission is a
\textit{quantum interference} (Fano) effect characterized by certain
discrete energy states interacting with the continuum. The \textit{width} of
the transmission line-shape is determined by the ratio of the effective
\textit{coupling} strength $G=gJ_{0}(\Omega /\nu )$ to group velocity $v_{g}$%
. Obviously, the decoupling of the TLS-photon interaction occurs when $%
J_{0}(\Omega /\nu )$ vanishes. Consequently, the transmission dip
disappears, i.e., the single photon goes through the TLS scatterer without
absorption or emission. In other words, \textit{by adjusting the amplitude
and frequency of the modulating field $\Omega _{a}(t)$, the TLS spontaneous
emission and stimulated transition are suppressed}. Measuring the photon
transmission or reflection would show the occurrence of the dynamic QZE.

\begin{figure}[ptb]
\includegraphics[bb=40 510 547 785, width=8 cm, clip]{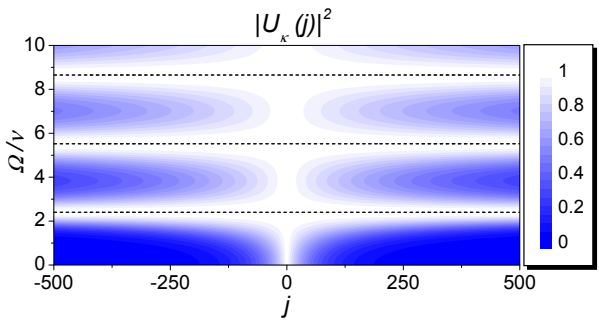}
\caption{(color online) Contour plot of the square $|U_{\protect\kappa%
}(j)|^{2}$ of the bound-state wave function amplitude versus the ratio $%
\Omega/\protect\nu$ and the space coordinate $j$. The parameters here are
the same as in Fig.~\protect\ref{fig:qdz-2}. The horizontal dashed lines are
the zeros of $J_{0}(x)$, corresponding to extended (along $j$) photon
wavefunctions}
\label{fig:qdz-3}
\end{figure}

The CRW with the TLS embedded in can also support localized modes. We now
discuss how to observe the dynamic QZE from the localization of photons,
described by the photonic bound state in the CRW. The bound state is assumed
to have the following solutions with even parity for the eigenvalue
equation~(\ref{eigeneq}):%
\begin{align}
U_{\kappa}\left( j\right) & =C\exp(-\kappa j),j>0)  \notag \\
U_{\kappa}\left( j\right) & =C\exp(\kappa j),j<0),
\end{align}
which is localized around the 0th site where the TLS is embedded. Here, the
imaginary wave vector $\kappa$ labels the energy
\begin{equation}
E_{\kappa}=\Delta \pm2\xi \cosh \kappa
\end{equation}
of a localized photon within the energy gap. The continuity of $U_{\kappa
}\left( j\right) $ at the boundary $j=0$, i.e. Eq.(\ref{5boundc}) determines
the existence condition of the bound state
\begin{equation}
g^{2}J_{0}^{2}\left( \frac{\Omega}{\nu}\right) =2\xi(\xi \sinh2\kappa \pm
\Delta \sinh \kappa).
\end{equation}
where the lower sign gives the living condition for the bound state which
lies upper the band with energy $E_{\kappa}=\Delta+2\xi \cosh \kappa$. On
the resonance $\Delta=0$, the width of the bound state is determined by%
\begin{equation}
\sinh2\kappa=\frac{g^{2}J_{0}^{2}(\Omega/\nu)}{2\xi^{2}},
\end{equation}
which can be adjusted by the ratio $\Omega/\nu$ through $J_{0}\! \left(
x\right) $. At the zeros of this Bessel function, $\kappa=0$ and the photon
is delocalized. When off resonance, the energies of the two bound states are
asymmetric with respect to the center of the band. In Fig.~\ref{fig:qdz-3},
we plot the single-photon distribution $|U_{\kappa}(j)|^{2}$ as a function
of the discrete coordinate $j$, and the ratio $\Omega/\nu$. It shows that:
(1) The wave packet of the single-photon spreads along the CRW as $%
\Omega/\nu $ increases, because%
\begin{equation*}
J_{0}\sim x^{-1/2};
\end{equation*}
(2) The width of the localized photon state oscillates following $%
J_{0}(\Omega/\nu)$; (3) Due to this modulation-induced Zeno effect, the wave
packet will be extended at the zeros of $J_{0}(x)$; (4) The full-spreading
of the photon wave packet does not occur as a periodic function of $%
\Omega/\nu$ because the roots of $J_{0}(x)$ are not periodic. The above
investigation indicates that the dynamic Zeno effect can be characterized by
the delocalization of photons.

\section{\label{Sec:6}Conclusion}

Based on the dynamical QZE induced by a high-frequency modulation, we show
how to decouple a TLS from its surroundings. We apply this QZE-based
decoupling mechanism to realize a quantum switch for a photon. The dynamic
QZE switch studied here allows to control the single-photon transport in a
CRW by using a TLS which is periodically modulated in time. Our analytical
results agree well with our numerical results using Floquet theory. This
proposal should be realizable experimentally~\cite{YouPT58}, because the
quantum Zeno switch only depends on the ratio between the amplitude $\Omega$
and the frequency $\nu$ of the external modulation. This QZE photon switch
might be useful to quantum information science, for the coherent control of
single photons.

We note that if one puts atoms in a MOT, with the atomic frequency modulated
by the AC Stark effect, and observes the light scattering, complete
reflection won't occur at the correct ratio of modulation strength to
frequency, which just like the one discussed here. However, due to the
linear dispersion of the reservoir which atoms in a MOT interacts with,
photon localization can not be observed.

As we know that, in reality, all quantum systems interact with the
environment, which results in the inelastic scattering of a single
photon. The inelastic scattering of photons would reduce the
transmission of photons as well as the quantum switching efficiency.
The decoherence or dissipation either influences the free
propagation of the single photon, or broadens the width of the line
shape at the resonance, according to its contributions to the
scattering process. It is known that photon propagates freely in
these resonators, all the dissipation factors of the resonators have
effect on the free propagation of the single photon. However, the
decay of the two-level system influences the scattering process,
since the energy of photon is not conservative before and after its
interacting with the two-level system, therefore, the width of the
lineshape is broadened.

This work is supported by New Century Excellent Talents in
University (NCET-08-0682), NSFC No.~10935010, No.~10474104, and
No.~10704023, NFRPC No.~2006CB921205, and 2007CB925204, and
Scientific Research Fund of Hunan Provincial Education Department
No.~09B063. FN acknowledges partial support from the NSA, LPS, ARO,
NSF grant No.~EIA-0130383, JSPS-RFBR 06-02-91200, and the JSPS-CTC
program. We thank S. Ashhab for useful discussions. \vspace*{-0.1in}

\appendix

\section{Quantum Zeno Dynamics of Two Level System in An Artificial Bath}

In this appendix we follows the references~\cite{KofPRL87} to discuss the
quantum zeno dynamics of the two level system in the coupled-resonator
waveguide (CRW), which behave as an artificial bath.

To this end, we make a Fourier transformation%
\begin{equation}
\hat{a}_{j}=\frac{1}{\sqrt{N}}\sum_{k}e^{ikj}\hat{a}_{k}
\end{equation}%
for Hamiltonian $H_{\mathrm{C}}+H_{\mathrm{I}}$ in Eqs.(\ref{lps-01}) and (%
\ref{lps-02}). In the subspace supported by the complete basis $\{\left\vert
e0\right\rangle ,\left\vert gk\right\rangle \}$, the Hamiltonian can be
rewritten as
\begin{equation}
H=\sum_{k}[\omega _{k}\left\vert k\right\rangle \left\langle k\right\vert +%
\frac{J}{\sqrt{N}}\left\vert kg\right\rangle \left\langle 0e\right\vert +%
\mathrm{H.c.}]+\Omega _{\mathrm{a}}(t)\left\vert e\right\rangle \left\langle
e\right\vert
\end{equation}%
where $\omega _{k}=\omega _{c}-2\xi \cos k$. Therefore, the state at
arbitary time
\begin{equation}
\left\vert \phi \left( t\right) \right\rangle =\sum_{j}u_{j}\!\left(
t\right) \left\vert 1_{j}\,g\right\rangle +u_{e}\!\left( t\right) \left\vert
0\,e\right\rangle
\end{equation}%
lies within the subspace spanned by $\left\vert 0e\right\rangle $ and $%
\left\vert kg\right\rangle $. Then we obtain the equations for amplitudes $%
u_{k}$ and $u_{e}$ by substituting the state $\left\vert \phi \left(
t\right) \right\rangle $ into the Schr\"{o}dinger equation
\begin{align}
i\partial _{t}U_{k}& =\frac{J}{\sqrt{N}}U_{e}\mathrm{e}^{-i\left( \omega
_{a}-\omega _{k}\right) t}\mathrm{e}^{-i\frac{A}{\upsilon }\sin \left(
\upsilon t\right) }  \notag \\
i\partial _{t}U_{e}& =\frac{J}{\sqrt{N}}\sum_{k}U_{k}\mathrm{e}^{i\left(
\omega _{a}-\omega _{k}\right) t}\mathrm{e}^{i\frac{A}{\upsilon }\sin \left(
\upsilon t\right) }
\end{align}%
where the relation between the capital letter and the lowercase is given by
\begin{equation}
u_{k}=U_{k}\mathrm{e}^{-i\left( \omega -2\xi \cos k\right)
t},u_{e}=U_{e}e^{-i\left[ \omega _{a}t+\frac{A}{\upsilon }\sin \left(
\upsilon t\right) \right] }
\end{equation}%
The above differential equations about $U_{k}$ and $U_{e}$ give an exact
integro-differential equation for $U_{e}\left( t\right) $
\begin{equation}
\dot{U}_{e}=-\epsilon ^{\ast }\left( t\right) \int_{0}^{t}d\tau U_{e}\left(
\tau \right) \epsilon \left( \tau \right) \Phi \left( t-\tau \right) \mathrm{%
e}^{i\Delta \left( t-\tau \right) }  \label{appdue}
\end{equation}%
where the detuning $\Delta =\omega _{a}-\omega _{c}$. Here
\begin{equation}
\Phi \left( t\right) =\frac{J^{2}}{N}\sum_{k}\mathrm{e}^{i2\xi t\cos k}
\end{equation}%
is the memory function or call reservoir response function, and the
modulation function reads
\begin{equation}
\epsilon \left( t\right) =\mathrm{e}^{-i\frac{A}{\upsilon }\sin \left(
\upsilon t\right) }\text{.}
\end{equation}

In the weak coupling limit, the approximation $U_{e}\left( \tau \right)
\approx U_{e}\left( t\right) $ can be made on the right-hand side of the
equation (\ref{appdue}). Then we have
\begin{widetext}
\begin{equation}
U_{e}\left(  t\right)  =\exp \left[  -\int_{0}^{t}dt_{1}\epsilon^{\ast}\left(
t_{1}\right)  \int_{0}^{t}dt_{2}\epsilon \left(  t_{2}\right)  \Phi \left(
t_{1}-t_{2}\right)  \mathrm{e}^{i\Delta \left(  t_{1}-t_{2}\right)  }%
\Theta \left(  t_{1}-t_{2}\right)  \right]
\end{equation}
\end {widetext}The Fourier transformation of $\Phi \left( t\right)
e^{i\Delta t}\Theta \left( t\right) $ is obtained
\begin{equation*}
\tilde{\Phi}\left( \omega +\Delta \right) /2=\frac{J^{2}}{2N}\sum_{k}\delta
\left( \omega +\Delta +2\xi \cos k\right) ,
\end{equation*}%
which is expressed in terms of the coupling spectrum density $\tilde{\Phi}%
\left( \omega \right) $. The coupling spectrum is defined as the Fourier
transformation of the reservoir response function
\begin{equation}
\tilde{\Phi}\left( \omega \right) =\frac{1}{2\pi }\int_{-\infty }^{\infty
}\Phi \left( t\right) \mathrm{e}^{i\omega t}dt.
\end{equation}%
Further introducing the the effective interaction time%
\begin{equation}
Q\left( t\right) =\int_{0}^{t}d\tau \left\vert \epsilon \left( \tau \right)
\right\vert ^{2}
\end{equation}%
and defining the normalized spectral modulation intensity%
\begin{eqnarray*}
F_{t}\left( \omega \right) &=&Q^{-1}\left( t\right)
\int_{0}^{t}dt_{1}\epsilon ^{\ast }\left( t_{1}\right) e^{-i\omega
t_{1}}\int_{0}^{t}dt_{2}\epsilon \left( t_{2}\right) e^{i\omega t_{2}} \\
&=&Q^{-1}\left( t\right) \left\vert \int_{0}^{t}dt^{\prime }\epsilon \left(
t^{\prime }\right) \mathrm{e}^{i\omega t^{\prime }}\right\vert ^{2},
\end{eqnarray*}%
the norm of amplitude reads%
\begin{equation*}
\left\vert U_{e}\left( t\right) \right\vert =\exp \left[ -\frac{Q\left(
t\right) }{2}\int_{-\infty }^{\infty }\tilde{\Phi}\left( \omega +\Delta
\right) F_{t}\left( \omega \right) d\omega \right]
\end{equation*}%
The effective decay rate
\begin{equation}
R\left( t\right) =\int_{-\infty }^{\infty }\tilde{\Phi}\left( \omega +\Delta
\right) F_{t}\left( \omega \right) d\omega  \label{appdr}
\end{equation}%
is written as the overlap of the reservoir coupling spectrum and normalized
spectral modulation intensity.

If the states $\left\vert k\right\rangle $ belongs to a spectrally dense
band, the CRW forms a `reservoir" spectral distribution, $\tilde{\Phi}\left(
\omega \right) =J^{2}\rho \left( \omega \right) $, with spectral density
given by
\begin{eqnarray*}
\rho \left( \omega \right) &=&\int_{-\pi }^{\pi }\delta \left( \omega +2\xi
\cos k\right) dk \\
&=&\frac{1}{4}\int_{-\infty }^{\infty }e^{-i\omega x}J_{0}\left( 2\xi
x\right) dx \\
&=&\left\{
\begin{array}{c}
0\text{ \ \ \ \ \ \ \ \ \ \ \ }2\xi <\left\vert \upsilon -\omega \right\vert
\\
\infty \text{ \ \ \ \ \ \ \ \ \ \ \ }2\xi =\left\vert \upsilon -\omega
\right\vert \\
\frac{1/2}{\sqrt{4\xi ^{2}-\omega ^{2}}}\text{ \ \ }2\xi >\left\vert
\upsilon -\omega \right\vert%
\end{array}%
\right.
\end{eqnarray*}

$R\left( t\right) $ in Eq.(\ref{appdr}) was obtained by Kofman and Kurizki
for very general cases to discuss the dynamics of Zeno and anti-Zeno effects
in a heat bath. We only demonstrate their universal approach with our setup
for coherent control of single photon transfer. The detailed calculation for
the decay rate begins with the effective interaction time $Q\left( t\right)
=t$ and the modulation spectrum%
\begin{eqnarray*}
F_{t}\left( \omega \right) &=&\sum_{n=-\infty }^{+\infty }J_{n}^{2}\left(
\frac{\Omega }{\upsilon }\right) \mathrm{sinc}^{2}\omega _{n}+ \\
&&\sum_{n_{1}\neq n_{2}}^{+\infty }tJ_{n_{1}}\left( \frac{\Omega }{\upsilon }%
\right) J_{n_{2}}\left( \frac{\Omega }{\upsilon }\right) \mathrm{sinc}%
^{2}\omega _{n_{1}}\mathrm{sinc}^{2}\omega _{n_{2}}
\end{eqnarray*}%
where $\omega _{n}=\left( \omega -n\upsilon \right) t/2$ and $\mathrm{sinc}%
x=\sin x/x$. It yeilds the effective decay rate
\begin{equation*}
R\left( t\right) =\sum_{k}\frac{tJ^{2}}{N}\left\vert \sum_{n=-\infty
}^{+\infty }J_{n}\left( \frac{A}{\upsilon }\right) \frac{\sin \frac{\left(
\Delta +2\xi \cos k+n\upsilon \right) t}{2}}{\left( \Delta +2\xi \cos
k+n\upsilon \right) t/2}\right\vert ^{2}.
\end{equation*}%
When no modulation is applied, i.e. $A=0$ and $t\rightarrow \infty $
\begin{equation*}
R\left( t\right) =2\pi \tilde{\Phi}\left( \Delta \right)
\end{equation*}%
which is the extension of the golden rule rate to the case of a
time-dependent coupling. In this paper, we are interested in a regime with
parameters $\Delta ,\xi \ll \nu $, which means that the modulation frequency
is much greater than the inverse correlation time of the continuum. The
effective decay rate reads%
\begin{align*}
R\left( t\right) & =\sum_{k}\frac{tJ^{2}}{N}\left\vert J_{0}\left( \frac{%
\Omega }{\upsilon }\right) \frac{\sin \frac{\left( \Delta +2\xi \cos
k\right) t}{2}}{\left( \Delta +2\xi \cos k\right) t/2}\right\vert ^{2} \\
& =tJ_{0}^{2}\left( \frac{\Omega }{\upsilon }\right) \int_{-\infty }^{\infty
}\tilde{\Phi}\left( \omega \right) \left\vert \frac{\sin \frac{\left( \omega
-\Delta \right) t}{2}}{\left( \omega -\Delta \right) t/2}\right\vert
^{2}d\omega
\end{align*}%
The above equation shows that the decay rate is determined by: 1) the
parameters of the driving field, i.e. ratio of modulation strength $\Omega $
to frequency $\upsilon $. 2) the overlap the reservoir coupling spectrum and
the modulation spectrum. Therefore the width and the center of the spectrum
are important factors. Obviously, the width of $\tilde{\Phi}\left( \omega
\right) $ is less than $4\xi $, the center of $\tilde{\Phi}\left( \omega
\right) $ lies in $\omega _{M}^{G}=0$. The width of $F_{t}\left( \omega
\right) $ is $t^{-1}$, $\Delta $ is its centra. This following results can
be obtained: 1). If ratio of modulation strength $\Omega $ to frequency $%
\upsilon $ satisfy $J_{0}\left( \Omega /\upsilon \right) =0$, then the
effective decay rate $R\left( t\right) =0$. 2). For sufficiently long times,
i.e. $t$ is much larger than both $\upsilon ^{-1}$ and an effective
correlation time $\left( 4\xi \right) ^{-1}$, the decay rate
\begin{equation}
R\left( t\right) =J_{0}^{2}\left( \frac{\Omega }{\upsilon }\right) \tilde{%
\Phi}\left( \Delta \right) ,
\end{equation}%
3). When time $t\sim \upsilon ^{-1}$, with $\upsilon \gg \Delta ,2\xi $, the
normalized spectral modulation intensity is a small varying function over
the interval $4\xi $, therefore one can make the approximation%
\begin{equation*}
R\left( t\right) =tJ_{0}^{2}\left( \frac{\Omega }{\upsilon }\right) \mathrm{%
sinc}^{2}\frac{\left( \omega -\Delta \right) t}{2}\int_{-\infty }^{\infty }%
\tilde{\Phi}\left( \omega \right) d\omega .
\end{equation*}%
Since the effective decay rate is averaged over all decay channels, the
quantum Zeno effect generally occurs.

\end{document}